\documentclass[pre,aps,onecolumn,showpacs,floatfix,superscriptaddress]{revtex4}
\usepackage{graphicx}% Include figure files
\usepackage{dcolumn}% Align table columns on decimal point
\usepackage{color}% color package
\usepackage{hyperref}% hyper link cite package
\usepackage{latexsym}
\usepackage{amsmath, amsthm, amssymb}
\usepackage{epsfig}
\usepackage{latexsym}
\usepackage{bm}
\begin{document}
\title{Fragmentation of a sheet by propagating, branching and merging cracks}
\author{Deepak Dhar}
\affiliation{\small{Tata Institute of Fundamental Research, Homi Bhabha Road, Mumbai 400005, India}}

\begin{abstract}
We consider a model of fragmentation of sheet by cracks that move with a velocity in preferred direction, but undergo random transverse displacements
as they move. There is a non-zero probability of crack-splitting, and the split cracks move independently. If two cracks meet, they merge, and move as a single crack. In the steady state, there is non-zero density of cracks, and the sheet left behind by the moving cracks is broken into a large number of fragments of different sizes.   The evolution operator for this model  reduces to the Hamiltonian of quantum XY spin chain, which is exactly integrable.  This allows us to determine  the steady state, and also  the distribution of  sizes of fragments.

\end{abstract}


\maketitle
\section{Introduction}

There has been huge body of work in engineering and physics literature dealing with the distribution of sizes of fragments, when  a large piece of solid is broken into much smaller fragments, driven by its applications in many industrial processes like mining and milling, in geophysics and environmental science \cite{grady_kipp,herrmann_book,turcotte,astrom_rev_2006}.  The theoretical approaches to modelling fragmentation have generally focussed on the distribution of sizes of fragments. It is expected that the general form of the distribution would be independent of material details, and may be captured in  a simple theoretical model based on some general  physical principles.  
In one of the first theoretical attempts in this direction,  Mott and Linfoot \cite{mott}  modelled the fragmentation  as caused by randomly drawn Poisson-distributed horizontal and vertical lines. Grady and Kipp studied different variants of the Mott construction, and their effect on the fragment-size distribution \cite{grady_kipp}. Similar ideas were also used by Gilvarry \cite{gilvarry1, gilvarry2} to predict the fragment size distribution. These have been studied
using the chemical reaction rate equations approach \cite{redner}. Several random fuse-network type  models for fracture have been developed, where the material is seen as  a network of coupled elastic elements, which have random threshold for failure \cite{arcangelis_redner,nakula} . There are also molecular-dynamic simulations   of assembly of sub-units, interacting by say Lennard-Jones interaction, subjected to high-impact projectile, or expansive stress \cite{herrmann_2012,meakin1,meakin2,anderson,astrom2000}. Models assuming a fractal structure of defects have also been studied \cite{perfect}.

\begin{figure}
\vspace{-0.4cm}
{\includegraphics[angle=90, width=0.6  \columnwidth]{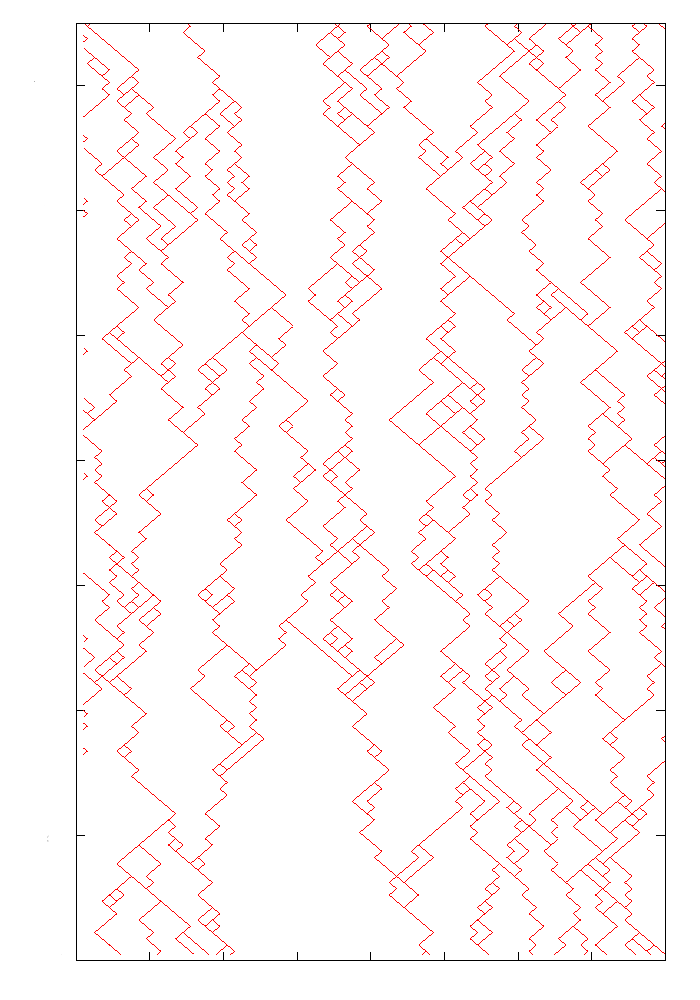}}
\caption{ Structure of the fragments formed in our Model A. The picture shown is obtained from a numerical simulation, size of lattice $L=80$, evolved to depth $150$, with  the crack- splitting rate  $\lambda = 0.1$. Note the wide distribution of fragment sizes.  }
\label{fig1}
\end{figure}

 In recent years, there has been a lot of concern about climate change, leading to increased melting  of the polar ice,  and consequent rise of sea-levels. To make quantitative predictions, one needs to understand the fragmentation of the the polar ice cap into much smaller iceberg  fragments. This led to recent study of calving glaciers by Astrom et al \cite{astrom1}.  The precise model studied by Astrom  et al is rather complicated, and has to be solved numerically \cite{astrom2, astrom3}. In this paper, we propose a simplified model of stochastic growth of cracks on a  two dimensional sheet, which was inspired by these papers. The cracks have a preferred direction of propagation, but have a random transverse velocity. There is a finite probability  per unit time that the a crack splits into two, which then propagate independently. If two cracks meet, they merge, and evolve subsequently as  a  single crack.   The region left behind by the cracks is broken into fragments of different sizes.  An example of the fragments generated in our model is shown in Fig. \ref{fig1}. 

This model was studied earlier by Ben Avraham et al \cite{benavraham, doering} as a model of particles with diffusion, aggregation and birth. They noted that the detailed balance condition is satisfied, and hence  the steady state is  easily written down, and  showed that the spacing distribution between consecutive particles satisfies a linear equation, and hence can be determined. The evolution operator for this model reduces to the Hamiltonian of a quantum XY spin model, which  well known to be diagonalizable exactly in terms of free fermions \cite{krebs, alcaraz,peschel, hinrichsen}.    Viewed as a model of fragmentation,  geometrical questions about  fragment-size distribution and correlations seem quite natural,  but these  have not been discussed before.  
A model  rather similar in spirit  was  discussed earlier by Inaoka and Takayasu \cite{inaoka_takayasu}, but   their detailed model is actually  quite different, and can only be solved numerically.   Our model has the advantage of been analytically tractable, and we  are able determine the exact distribution of fragment sizes. We show that in the limit when the  splitting rate is small,   there is an exact mapping between our  model A (defined precisely below) and the directed abelian sandpile model \cite{ddrr,dd2}, which brings out  the  self-organized origin of the  divergence in the fragment-size distribution  for small sizes. In fact, this distribution is the same as that encountered in a rather different model, in which particles aggregate, not undergo fragmentation, also  studied by Takayasu  earlier\cite{takayasu_aggr}.

\section{Definition of the model}

We start  with   an intact cylinder  with  no cracks.  At the start, we introduce one or more cracks at the left boundary. We can define  a discrete-time update rule  (Model A), or continuous-time update rule ( Model B).\\
~\\

\begin{figure}
\vspace{-0.4cm}
{\includegraphics[angle=0, width=0.6  \columnwidth]{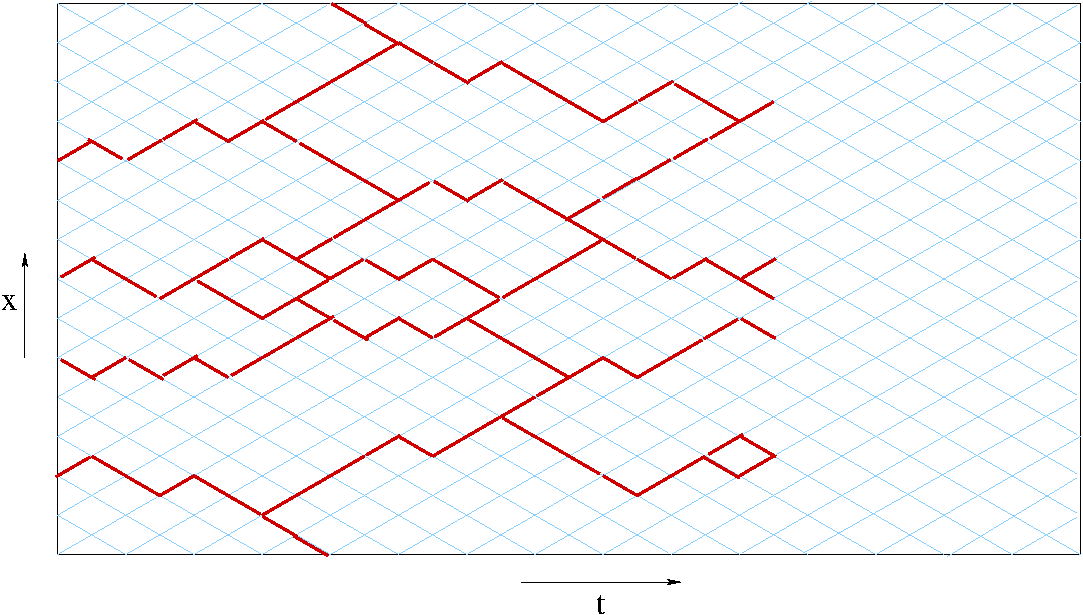}}
\caption{ A system of cracks propagating into a plane leaving behind a sea of fragments in our Model A. }
\label{fig2}
\end{figure}

{\bf Model A}: We consider the model defined on a square grid, drawn on the surface of  semi-infinite cylinder. The sites of the lattice are labelled by integers $(x,t)$, with $1 \leq x \leq 2L$, $t \geq 0$, and $(x+t)$ even.  The neighbours of site $(x,t)$ are four  the sites $ (x \pm 1,t \pm 1)$.  The periodic boundary conditions along the cylinder are imposed by identifying $(X + 2L,t) \equiv (x,t)$, for all $x,t$. This corresponds to the axis of the cylinder being in the direction $(1,1)$ of the lattice ( fig. \ref{fig2}).

 We consider discrete-time parallel update, where each crack moves inwards with speed $1$. Then, the cracks reach up to distance $t$ along the cylinder axis, and we may equivalently think of $t$ as the time coordinate.  If there is a crack at site $(x,t)$, it moves according to  the following rule:\\
a) With probability $\lambda$, it splits into two cracks,  and these cracks go   to $(x+1,t+1)$ and $(x-1,t+1)$. Clearly, $0 \leq \lambda \leq 1$.\\
b) If the crack does not split, it moves to any one of the two forward neighbours $(x+1,t+1)$ or $(x-1,t+1)$ with equal probability.\\
c) If two cracks reach a site, they merge, and move as a single crack. \\
~\\

{\bf Model B}: Here we consider  the time-coordinate $t$  to be a  continuous variable, $ t> 0$, but the spatial coordinate is still discrete, $0 < x \leq 2L $, with periodic boundary conditions $ x+ 2 L \equiv x$. A configuration at time $t$ is specified by the occupation numbers $\{n_x\}$, where $n_x$ takes values $1$ and $0$, depending on whether there is a crack at $x$ at time $t$ or not. 

The configurations $\{n_x\}$ undergo continuous-time Markovian evolution with the following rates: \\
a) A crack at position $x$ jumps to  site $x'$ with rate $1$, if $x'$ is a nearest neighbour of $x$.\\
b) An empty neighbour $x'$ of an occupied site $x$  becomes occupied with  rate $\lambda$, due to crack-splitting at $x$.\\
c) If a crack arrives at a site already occupied by a crack, it merges with it, and the state of the site does not change.  

If we represent the configuration by a bit string, e.g.  $...00101001110001..$, the transition rates are given by the equations

\begin{eqnarray}
~~~ 01  \leftrightharpoons  10, ~~~ {\rm~~~with~ rate~ 1};\\
~~~ 01  \rightarrow 11, ~~~ 10 \rightarrow 11, {\rm ~~with~rate~ \lambda};\\
~~~ 11  \rightarrow 01, ~~~ 11 \rightarrow 10, {\rm ~~with~rate~ 1}.
\nonumber
\end{eqnarray}

\begin{figure}
\vspace{-0.4cm}
{\includegraphics[angle=0, width=0.6  \columnwidth]{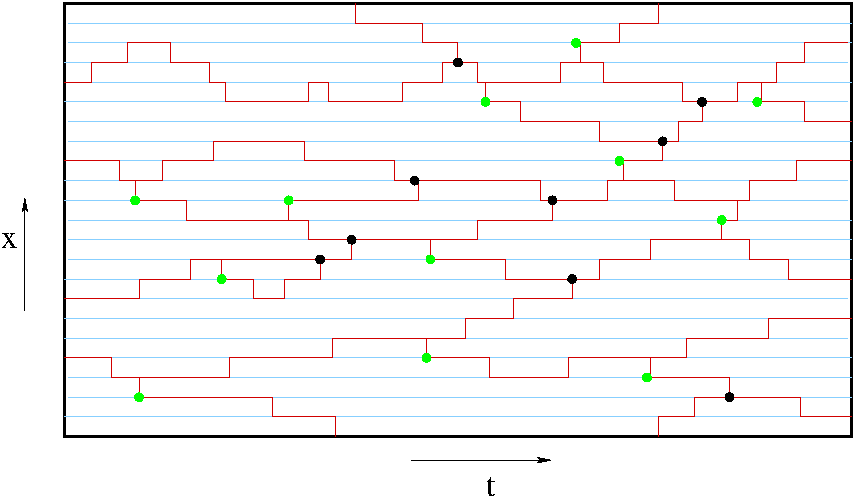}}
\caption{ An example of the fragments of the plane left behind in model B, after  the crack front  has advanced beyond the region shown. Here time is continuous.  The blue and red lines denote world lines of  sites, and of cracks respectively. The green dots denotes crack creation, and black dots denote crack mergers.  }
\label{fig3}
\end{figure}

Fig. \ref{fig3} shows a schematic time-history of evolution in this case.

\section{Calculation of the steady state}
In our model, the number of cracks at any given time ( equivalently at distance along the axis of the cylinder) is not conserved. So long as at the start of the configuration $t=0$, there is at least one crack, as the cracks propagate along the cylinder, they may split. or merge, and for large times, there is a non-trivial steady state of the system with a non-zero density  $\rho(\lambda)$ of cracks, and a finite mean area per fragment. We now determine the this steady state, and the asymptotic density.\\
~\\

{\bf Model A}: Let us label the   bonds of the lattice by coordinates of their midpoints.  Then a bond has coordinates $(m+1/2, n+1/2)$, where $m$ and $n$ are any integers, with $0 \leq m \leq M-1$, and $ 0 \leq n < \infty$.  The configuration ${\bf \Sigma}$ of cracks is specified by giving the occupation numbers $\{ \sigma(m,n)\}$ of all the bonds, where $\sigma(m,n) =1$, if the bond $(m+1/2, n+1/2)$ is occupied by a crack, and zero otherwise. The restriction of ${\bf \Sigma}$ to vertical row $t$ will be denoted by ${\bf \Sigma}_t$

We can think of the set $\{ \sigma(m,n) \}$ as the evolution history of a set of  $2M$  Ising spins on a line, undergoing stochastic Markovian dynamics.
Here $\sigma(m,n)$ is  $+1$ if the m-th spin  is up at time $t$, and zero otherwise.  In this formulation, the Markovian evolution is defined as follows: At odd times, we consider pairs of adjacent spins $[\sigma(2j,t),\sigma(2j+1,t)]$. These evolve according to the following rules:\\
a)  $[0,0]$  remains unchanged.\\
b)  If $[\sigma, \sigma'] \neq [0,0]$, it is reset to $(1,0), (0,1)$ or $(1,1)$ with probabilities $(1 -\lambda)/2,  (1 -\lambda)/2, \lambda$ respectively.

At even times $t$, we apply the same rule, but choose the pairs to be $[\sigma(2j+1,t), \sigma(2j+2,t)]$. 

Now, we  specify the state of the system at time $t$, by giving a probability measure on the space of spin configurations. Let ${\rm Prob}({\bf \Sigma}_t)$ denote the probability that the configuration of Ising spins  at time  $t$ will be found to be  ${\bf \Sigma}_t$. Note that there are $(2^{2L}-1)$ allowed values of ${\bf \Sigma}_t$, as the configuration with all spins down is not reached in the steady state, if we start with at least one crack present at $t =0$. The Master equation for the evolution of $ {\rm Prob}({\bf \Sigma}_t)$ is
\begin{equation}
{\rm Prob}({\bf \Sigma}_{t+1}) = {\mathbb W}[{\bf \Sigma}_{t+1}, {\bf \Sigma}_t]~~ {\rm Prob}({\bf \Sigma}_t)
\end{equation}
where ${\mathbb W}$ is the Markov transition matrix. 

We note that the transition rates here satisfy the detailed balance condition corresponding to the Hamiltonian
\begin{equation}
{\mathbb H}({\bf \sigma}) = -K \sum_{i=1}^{M}  \sigma_i
\label{eqH}
\end{equation}
where $e^K  = 2 \lambda/(1-\lambda)$.

There are $2^{2L}$ states, and ${\mathbb W}$ is a $4^L \times 4^L$ matrix. There are two steady state eigenvectors: one in which there are no cracks, and one in which a configuration with $n$ cracks has a probability proportional to $e^{Kn}$.

{\bf Model B}: In this case, the analysis is simpler.  We note that probability  of transition  of the state of two adjacent sites from the state  $01$ to $11$ is $\lambda $, but the rate of transition from $11$ to $01$ is $1$.  Also, there is a rate $1$ for the transition $01$ to $10$, and vice-versa. 
Then, detailed balance is clearly satisfied, for  the Hamiltonian function given by Eq.(\ref{eqH}), with $\lambda = e^K$.

\section{Distribution of fragment sizes}

We first discuss  the Model A. The cracks divide the plane into fragments of different sizes. We note that each fragment is a polygon having a unique starting point, and a unique end-point which are the boundary sites with minimum and maximum values of the time-coordinate respectively. The boundary itself consists of two directed walks. These shapes are known as staircase polygons in literature. 

The left boundary of the fragment  is a simple directed biassed   random walk, where the step occurs to the right with probability $(1 +\lambda)/2$, and to the left with probability $(1 -\lambda)/2$. Similarly, the right boundary is an independent  walk with step to the left with probability $(1 +\lambda)/2$, and step to the right occurs with probability $(1-\lambda)/2$. The fragment ends when these two boundaries meet.

Consider the stochastic evolution of the system.  Suppose there is a tip-splitting event at some site $(x_0,t_0)$, which starts  a new fragment. Let ${\rm Prob}(A, \lambda)$ be the probability that this fragment will have area $A$.  For small $A$, these are easily determined by explicit enumeration of possible shapes. For example,  consider a fragment starting at $(x_0,t_0)$. We can have $A=1$, only if  the fragment boundary, when at the site $(x_0+1,t_0+1)$ takes the next step to the  left, and the boundary, when at $(x_0-1,t_0+1) $ takes the next step to the right. Thus we have ${\rm Prob}(A=1, \lambda) = (1+ \lambda)^2/4$.   Similarly, ${\rm Prob}(A=2, \lambda)= ( 1 + \lambda)^2 (1 - \lambda^2)/8$.  More generally, if a given fragment shape ${\mathcal S}$ has perimeter $2P$, it occurs with a probability $ ( 1 + \lambda)^2 (1 - \lambda^2)^{P-1} 2^{-P-1}$. 

Thus, the question of finding ${\rm Prob}(A)$ for general $A$ is reduced to the finding the number of different staircase polygons of area $A$.  This has studied earlier by Prellberg\cite{prellberg1}, and Prellberg and Brak \cite{prellberg2}. Let the number of  a staircase polygon having $2 h$ horizontal bonds, and $2 v$ vertical bonds, and area $A$ be $C(h,v,A)$.  We define the generating function
\begin{equation}
G(x,y,A) = \sum_{h, v, A} x^{h} y^{v} q^A
\end{equation}

Prellberg showed That $G(x,y,A)$ satisfies a functional equation, and  used it to determine $G(x,y,q)$ exactly. One gets
\begin{equation}
G(x,y,q) = y \left[ \frac{H(q^2x,qy,q)}{H(qx,y,q)}  -1 \right]
\end{equation}
where
\begin{equation}
H(x,y,q) = \sum_{n=0}^{\infty} \frac{ (-x)^n q^{n \choose 2}}{(q;q)_n (y;q)_n}
\end{equation}
with
\begin{equation}
(r;q)_n = \prod_{m=0}^{\infty} (1 - t q^m)
\end{equation}

These explicit formulas are a bit complicated.  These simplify in the continuum limit, corresponding to   $\lambda \rightarrow 0$. The  continuum  problem of  determining the probability distribution of area in a Brownian bridge has been studied in  \cite{kearney, kearney2}. They showed that in this case,  the distribution function ${\rm Prob}(A,\lambda)$ is the Airy distribution. 
For small $\lambda$,  ${\rm Prob}(A,\lambda)$ has the scaling form
\begin{equation}
{\rm Prob}(A,\lambda) \simeq  \lambda^4 g(\lambda ^3 A).
\label{eq:g}
\end{equation}

In the limit $\lambda \rightarrow 0$,  and fixed $A$, ${\rm Prob}(A,\lambda)$ tends to a non-zero value.  In this case,  the boundaries of the fragment do unbiased random walk,and the corresponding probabilities are exactly as found in the  directed abelian sandpile model in 2-dimensions \cite{ddrr}.  For this case, simple dimensional arguments already show that ${\rm Prob}(A) \sim A^{-4/3}$ \cite{ddrr}. This implies that
\begin{equation}
{\rm Prob}(A, \lambda~\rightarrow ~0) = {\rm Prob} (A)_{ASM}
\end{equation}
where ${\rm Prob} (A)_{ASM}$ is the probability that adding  a grain in the steady state of 2-dimensional directed abelian  sandpile model will cause an avalanche with exactly $A$ topplings \cite{ddrr,dd2}.  It is known that ${\rm Prob} (A)_{ASM} \sim A^{-4/3}$ for large $A$, we must have
\begin{equation}
g(x) \sim K x^{-4/3},{ \rm for ~ small~~x}.
\end{equation}
For large $x$,   the Airy distribution, $g(x)$ decreases as  $\exp( - x^{1/2})$ \cite{kearney2}.  In the literature, there has been some discussion about the size-distribution having exponential or stretched exponential decay  for large sizes \cite{grady_kipp}.

Note the unusual $\lambda^4$ scaling of the prefactor in Eq.(\ref{eq:g}).  This arises because the function $g(x)$, which varies as $x^{-4/3}$ for small $x$ is not integrable. Then, if we take the scaling limit 
\begin{equation}
\sum_{A=1}^{\infty} {\rm Prob}(A,\lambda) \simeq \int_{x_{min}}^{\infty} dx g(x),
\end{equation}
we have to impose a  cut-off  of $x_{min} = {\mathcal O}(\lambda^3) $ on the lower limit of the integral. Then the integral diverges as $x_{min}^{-1/3}$, which cancels the extra power of $\lambda $ in the Eq. (\ref{eq:g}). 

 For finite but small $\lambda$, the mean density of cracks varies as $1/\lambda$, and the average longitudinal extent of  a fragment scales as $1/\lambda^2$. Thus, the mean size of a fragment varies as $1/\lambda^3$.  

Now, we consider Model B. Let ${\rm Prob}(A,m) dA $ be the probability that a segment of width $m$ will generate a fragment with additional area between $A$ and $A + dA$.  The Laplace transform of this distribution will be denoted by $\tilde{P}(s,m)$, and is defined by
\begin{equation}
\tilde{P}(s,m) = \int_0^{\infty} dA ~~{\rm Prob}(A,m) e^{- sA}
\end{equation}

Consider the time evolution of a fragment whose transverse size is $m$ at some time $t=t_0$.  In time, the boundaries of this fragment will evolve stochastically, till they merge. 
The segment of length $m$ can become segment of length $ m \pm 1$, if either end makes a diffusive jump ( with rate $2$ for each), or if either end creates a new crack which decreases the size of this segment (  total rate $2 \lambda$). The waiting time $t_1$ before any one of these occurs is an exponentially distributed random variable. The  probability that this occurs between time $t_1$ and $t_1 +dt_1$ given by  ${\rm Prob}(t_1) dt_1 = (4+ 2 \lambda) e^{-(4 +2 \lambda)t_1} dt_1$.  The total area of the fragment is sum of two independent random variables: the area $A_1 =  m t_1 $ up to this time, and remainder area $A_2$.   The remainder has starting length  $m+1$ with probability $p_{_+} = 1/(2 +\lambda)$, and $m-1$, with probability $p_{_-} =  ( 1 + \lambda)/(2 + \lambda)$.  This gives
\begin{equation}
\tilde{P}(s,m) = \left[ \int_0^{\infty} Prob(t_1) dt_1 e^{ -s m t_1} \right] \left[ p_{_+}~ \tilde{P}(s,m+1) + p_{_-}~ \tilde{P}(s,m-1)\right]
\end{equation}
which simplifies to 
\begin{equation}
(4 + 2 \lambda + m s) \tilde{P}(s,m) =  2 \tilde{P}(s,m+1) + (2 + 2 \lambda ) \tilde{P}(s,m-1)
\end{equation}

Write $\tilde{P}(sm) = ( 1 + \lambda)^{-s/2} Q(s,m)$. Then $Q(s,m)$ satisfies the equation of the form
\begin{equation}
(\frac{4+ 2 \lambda + m s }{\sqrt{1 + \lambda}} -2  )Q(s,m) = \left[ Q(s,m+1) +  Q(s,m-1) -2  Q(s,m)\right]
\end{equation}
which is a discrete variant of the Airy equation.  In the continuum limit of small $\lambda$ ,  it reduces to the Airy equation, and the scaling form of the fragment distribution is the same as studied in  \cite{kearney}. 

\section{Ultra-locality of correlations between fragments}

This model has the remarkable property that  two  fragment starting at any two sites $(x_1,t_1)$  and $(x_2,t_2)$ are completely uncorrelated, unless they are actually touching each other.  This is true not only for their area, but also their shapes.  This lack of correlations between fragments is very surprizing, in view of the fact that the condition that the arrangement of fragments  has to tile  the plane imposes very strong geometrical constraints on their shapes.  The probability that a fragment starting at $(x_1,t_1)$  will have a given have any  given shape ${\mathcal S}$,   depends only on ${\mathcal S}$, and is completely  independent of the state of  other sites at  time $t_1$.  If there is another crack that comes near this crack, at best it can merge with this fragment's boundary, but this will not influence its subsequent motion.

We have hitherto considered all cracks moving into the unfragmented region with unit speed. This assumption is not really necessary. At the start of the simulation, we can assign to each site a random variable, taking one of three possible possible values,  which tells what would happen if a crack reaches that site: will it split into two, or go left unsplit, or go right unsplit.  Then, we can start with any configuration of cracks at the boundary on the left, and evolve cracks in any order, and the final configuration is independent of the order in which the cracks are updated, and the waiting time between these updates. 

This kind of ``ultra-locality" of cluster shapes is  also seen in percolation models, where the shapes of two percolation clusters are perfectly uncorrelated, unless they have a common boundary. The  present model may also be seen as a model of this type.  This property is much less obvious, if we look at the transfer matrix of the model.  Consider the configuration of a particular column  $t=t_0$.  We assume that a crack-splitting occurs at a point $(x_0,t_0)$.  The probability that the generated fargment with $(x_0, t_0)$ as its left-most point has area $A$, denoted by  ${\rm Prob}(A, \lambda)$ above, would in general be expressible in terms  the eigenvalues and eigenvectors of the transfer matrix, and matrix elements of some projection operators between eigenvectors. That this probability does not depend on the starting vector, so long as a splitting occurs at $(x_0,t_0)$, 
requires the transfer matrix to have very special structure.  For example, the transfer matrix ${\mathbb W}$ can be written as
\begin{equation}
{\mathbb W} = {\mathbb T}_1 {\mathbb T}_2
\end{equation}
where ${\mathbb T}_1$ and ${\mathbb T}_2$ are the matrices cooresponding to evolution at odd and even times.  But ${\mathbb T}_2$ can be written as 
a direct product of $L$ $4 \times 4$ matrices, where each matrix gives the transition probabilities from the state of a pair of adjacent sites 
$\sigma(2j-1,t), \sigma(2j,t)$ to the state $\sigma(2j-1,t+1), \sigma(2j,t+1)$.  Each of these is the configuration basis is easily seen to be of the form\\
\begin{center}
$ \left[\begin{array}{cccc}
1 & 0 & 0 & 0 \\
0 & a & a & a \\
0 & a & a & a \\
0 & b & b & b \end{array} \right]$
\end{center}
 
where $a = (1-\lambda)/2, b= \lambda$.  Clearly, the rank of ${\mathbb T}_2$, and hence of ${\mathbb W}$ is at most $2^{L}$.   Out of the $4^L$ eigenvalues of $L$, at least $4^L - 2^L$ are exactly zero. The large number of 
zero eigenvalues provides the  mathematical mechanism underlying the  the ultra-locality in this model.

I thank S. N. Majumdar for discussions, and for pointing out ref. \cite{benavraham} to me. This work is supported in part by  the  Department of Science and Technology, Government of India
via grant  DST-SR/S2/JCB-24/2005.

\date{\today}

\end{document}